\documentclass[11pt, showkeys ]{article}

\usepackage{authblk}
 \usepackage[a4paper, total={6.5in, 10.3in}]{geometry}
  \usepackage{graphics,epsfig}
\usepackage{epstopdf}
\usepackage{graphicx}
 
\usepackage{amsmath}
\usepackage{amssymb}
 \usepackage{hyperref}
\begin{document}
%opening
\title{ Ay\'{o}n--Beato--Garc\'{i}a  black hole coupled with a cloud of strings: thermodynamics, shadows and quasinormal modes }

\author[a]{Amit Kumar\footnote{ammiphy007@gmail.com}}
\author[a]{Dharm Veer Singh\footnote{veerdsingh@gmail.com} \footnote{Visiting Associate, IUCAA, Pune, India}}
\author[b,c]{Sudhaker Upadhyay\footnote{sudhakerupadhyay@gmail.com}\footnote{Corresponding author} \footnote{Visiting Associate at IUCAA Pune, India}}

\affil[a]{Department Physics,
Institute of Applied Science and Humanities, GLA University, Mathura - 281406, Uttar Pradesh, 
India }

\affil[b]{Department of Physics, K.L.S. College, Nawada, Magadh University, Bodh-Gaya, Nawada 805110, India}
 
\affil[c]{School of Physics, Damghan University, P.O. Box 3671641167, Damghan, Iran}
\date{}
\maketitle
\begin{abstract}
 We find an exact black hole solution for the Einstein gravity in the presence of Ay\'{o}n--Beato-–Garc\'{i}a non-linear electrodynamics and a cloud of strings. The resulting black hole solution is singular, and the solution becomes non-singular when gravity is coupled with  Ay\'{o}n--Beato-–Garc\'{i}a non-linear electrodynamics only. This solution interpolates between Ay\'{o}n--Beato-–Garc\'{i}a black hole, Letelier black hole and Schwarzschild black hole { in the absence of cloud of strings parameter, magnetic monopole charge and both of them, respectively}.  We also discuss the thermal properties of this black hole and find that the solution 
follows the modified first law of black hole thermodynamics. Furthermore, we 
estimate the solution's black hole shadow and quasinormal modes. 
 
\end{abstract}

\section{Introduction}\label{sec1}
 Black holes (BHs) are one of the fascinating predictions of general relativity. 
 This is one of the solutions of Einstein's field equation. 
 Several observations were performed to confirm its existence.
 The existence was first approved when  LIGO  detected the first gravitational waves from the BH  merger \cite{1}.
Recently, the Event Horizon Telescope  (EHT) detected the shadow of the BH at the centre of Messier $87^*$ galaxy \cite{2}.
 Recently,  EHT  observed the
shadow  of Sagittarius  $A^*$ BH  around the centre of Milky Way \cite{sg}. When the bright light comes in the vicinity of a BH, the light bends and is pulled by the intense gravity of the BH, creating a shadow around the BH  where no light can reach an observer.

 Born and Infeld were the first to introduce non-linear electrodynamics (NLED)  to eliminate the central singularity of a point charge and energy divergence \cite{ned}. 
 The NLED became more popular when   NLED emerged as limiting cases of specific string theory \cite{wit}.  It is well-known that the linearity of electrodynamics does not hold at high energies because of the influence of other (physical) fields.
 In these circumstances, NLED theories may be considered a suitable alternative.  NLED, as a  source of gravity, is
capable to create various new BH solutions \cite{sud01,sud02,sud03,das}. 
 Beyond these,  NLED theories play an important role in cosmology \cite{cos1,cos2,cos3}     and string theories \cite{fr, wit}.
  Ay\'{o}n--Beato-–Garc\'{i}a (ABG) proposed 
  the first BH with an NLED field satisfying the weak energy condition \cite{ay1}.
  Later, a new regular exact BH solution was obtained for an NLED coupled to Einstein's gravity \cite{ay2}.  Further, the magnetic stable BH solution to Einstein equations coupled to ABG NLED was proposed \cite{ay3}. In 2005,  four parametric regular   BH solutions were studied for the Einstein equations coupled to ABG NLED \cite{ay4} {and other regular BH solutions are given in \cite{Singh:2022xgi, Singh:2017qur, Filho:2023voz, Biswas:2022qyl, KumarWalia:2022aop, Belhaj:2022qmn, Kumar:2023cmo}.}
  
Letelier \cite{xu19} first proposed the cloud of strings (CS), a model for a pressure-less perfect fluid. Being a model for fundamental constituents of the universe, the strings may have essential implications from the astrophysical and cosmological
 points of view.
The CS has been utilised as the possible material source for the 
  Einstein equations and a generalised BH solution were found \cite{lt1,lt2, Singh:2020nwo}. Thus, studying space-time geometry in the presence of ABG NLED and CS will be interesting. Precisely, this is the motivation of the present investigation.

 The size and shape of the BH shadow may give crucial information regarding the BH parameters like mass,  spin and geometry \cite{roh, ped, Vishvakarma:2023tnl, Vishvakarma:2023csw, Zubair:2023cep}. 
For instance, a slightly distorted circular shadow 
 of a BH may provide information on the { fast rotating rate of  BH}.
 The study of the observed shape of the shadow may boost the understanding of both gravity in extreme conditions and 
BHs. Recently, Shadow of the   Reissner-Nordstr\"om  BH \cite{ss}, charged
massive BTZ BH \cite{ss1}, and BH   \cite{ss2} {coupled with  
NLED have been studied}. We know that BH shadow is closely related to the specific complex 
characteristic frequencies \cite{xx}  known as quasinormal modes (QNMs) \cite{qnm}. { The QNMs for regular BHs are studied in \cite{Mou22,Cai:2021ele,Gogoi:2021cbp,Murshid:2022ssj,Konoplya:2023ahd,Gogoi:2023ffh,Jafarzade:2021umv} and QNMs in Eikonal limit in 
\cite{Ladino:2023zqc,Li21}. Shadow and QNMs are studied in \cite{k20}. }

{The    BH thermodynamics {was} studied originally by Bekenstein \cite{11,bak} and Hawking \cite{bak0} by establishing a connection between entropy and the BH horizon area.  
Further, it is found that the entropy of the BHs is proportional to the area of horizon \cite{str,car}. 
{The}  BH thermodynamics has found interest in many ways  \cite{sud2, sud3,jy,bss,sud1, behn,jhap,jhap1,beh1}.
For example, {the} effects of $\alpha'$ corrections on the thermodynamics of a Reissner-Nordstr\"om BH is studied \cite{beh2}. The non-perturbative correction to the Horava-Lifshitz BH thermodynamics is also analysed \cite{beh3}.
Thermal analysis of BH in de Rham--Gabadadze--Tolley massive gravity in Barrow entropy framework {has been} done recently \cite{sud7}.
 
 In this paper, we derive an exact BH solution for the Einstein gravity in the background of ABG non-linear electrodynamics and a cloud of strings. Interestingly, the resulting BH solution is singular as the regularity due to non-linear electrodynamics is compensated by the cloud of strings. This solution interpolates between ABG {BH}, Letelier {BH} and Schwarzschild {BH  in the absence of magnetic monopole (MM) charge and both of them, respectively}. We explore the horizon structure of the obtained BH solution.  We also discuss the thermal properties of this BH solution and find that this
follows the modified first law of BH thermodynamics. Furthermore, we discuss the   BH shadow and quasinormal modes.}

The paper is outlined in the following manner. We deduce a new BH solution for the 
Einstein gravity in the background of the ABG NLED and CS in Sec. \ref{sec2}. In Sec. \ref{sec3}, we calculate the 
thermodynamics of the resulting BH solution and 
find that this follows a modified first-law of  BH thermodynamics.
In Sec. \ref{sec4}, we first calculate the photon radius 
{ by} solving the geodesic equation leading to shadow radius. We also estimate the QNMs for this BH solution.
Finally, we summarise the results of the paper in the last section. 
%%%%%%%%%%%%%%%%%%%%%%%%%%%%%%%%%%%%%%%%%%%%%%%%%%%%%%%%%%%%%%%%%%%%%%%%%%%%%%%%%%%%%%
\section{Exact solution of ABG  BH with CS}\label{sec2}
In this section, we construct a new BH solution in the presence of
 ABG NLED and CS. In this regard, we first write about Einstein's action 
for the gravity model coupled with  ABG NLED and CS as
\begin{equation}
S=\int d^4x \sqrt{-\mathrm{g}}\left[R+{\cal L}_{ABG}(F) \right]+S_{\text{CS}},
\label{action}
\end{equation}
where $\mathrm{g}$  and $R$  are the determinant of metric and  curvature scalar respectively. Here,  ${\cal L}_{ABG}(F) $ refers to Lagrangian density for  ABG  expressed in terms of invariant $F=\frac{1}{4}F_{\mu\nu}F^{\mu\nu}$, where $F_{\mu\nu}=\nabla_\mu A_\nu-\nabla_\nu A_\mu$ is the Maxwell field-strength 
tensor,  having following form \cite{ay1,ay2,ay3,ay4}:
\begin{equation}
{{{\cal L}_{ABG}(F)}}= \frac{F(1-3\sqrt{2g^2F})}{(1+\sqrt{2g^2F})^{3}}-\frac{3M}{g^3}\left(\frac{(2g^2F)^{5/4}}{(1+\sqrt{2g^2F})^{5/2}}\right),
\label{nonl1}
\end{equation}
{ where parameters  $M$ and $g$  are related with the mass and MM charge of the BH, respectively  \cite{ay1,ay2}. The electromagnetic field tensor $F_{\mu\nu}$ in $4D$ spherically spacetime is given by
\begin{equation} 
F_{\mu\nu}=2\delta^{\theta}_{[\mu}\delta^{\phi}_{\nu]}Z(r,\theta)=2\delta^{\theta}_{[\mu}\delta^{\phi}_{\nu]}\,g(r)\sin{\theta}{,}
\end{equation}
which leads to $g(r)=g$, where we choose the integration constant as $g$ and it is identified as the MM charge. \begin{equation}\label{7}
\frac{1}{4\pi}\int_{s^\infty} F ds=\frac{g}{4\pi}\int_0^{\pi}\int_0^{2\pi}\sin\theta \,d\theta\, d\phi=g,
\end{equation}
 where $s^{\infty}$ is the spherical surface at infinity.
 Using (\ref{7}), the field strength tensor can be simplified to
\begin{equation}
F_{\theta\phi}= g\sin\theta, \qquad F=\frac{1}{2}\frac{g^2}{r^4}.
\label{emt1}
\end{equation}

}

The explicit expression for the CS action as a source, described by the Nambu-Goto term,   is given by \cite{das}
\begin{equation}  
S_{\text{CS}}= \int_{\Omega}  \; m \sqrt{-\gamma}  d\lambda^{0} d\lambda^{1},
\end{equation}
{ where $m$ is the mass of each  string, $\gamma$ is the determinant of 
$\gamma_{ab}= \mathrm{g}_{\mu\nu}\frac{\partial x^{\mu}}{\partial \lambda^{a}} \frac{\partial x^{\nu}}{\partial \lambda^{b}}$.  $x^{\mu}=x^{\mu}(\lambda^a)$ is describe the string world sheet and $\lambda^a\equiv (\lambda^0,\lambda^1)$, where  $\lambda^{0}$  {and}  $\lambda^{1} $, respectively, are  a time-like and space-like coordinates \cite{24}.} The  $\Omega^{\mu \nu}$ is a bi-vector defined as
\begin{equation}
\label{eq:bivector}
\Omega^{\mu \nu} = \epsilon^{a b} \frac{\partial x^{\mu}}{\partial \lambda^{a}} \frac{\partial x^{\nu}}{\partial \lambda^{b}},
\end{equation}
{ where $\epsilon^{ab}$ refers to the two dimensional Levi-Civita tensor defined as $\epsilon^{01}=-\epsilon^{1 0}=1$. Furthermore, since $T_{\mu\nu}=2\partial L/\partial \mathrm{g}^{\mu\nu}$, and then $\partial_\mu(\sqrt{-\mathrm{g}}\rho \Omega^{\mu \nu})=0$. Here, $\rho$ is the density, which describes the case of a {CS} \cite{lt2}, and the $ \Omega^{\mu \nu}$  is the function of radial distance. The non-vanishing component of  $ \Omega^{\mu \nu}$ is  $ \Omega^{t r}= \Omega^{rt}$. Thus the energy-momentum tensor becomes $T^t_t=T^r_r=-\rho \Omega^{t r}$, and using the $\partial_t(r^2\Omega^{t r})=0$  \cite{lt2}
\begin{equation}
T^t_t=T^r_r=\frac{a}{r^2},
\end{equation}
 {where $a, 0<a<1$ is an integration constant related to strings (so-called CS parameter}).
}
Varying the action (\ref{action}) concerning  $\mathrm{g}_{\mu\nu}$  and  $A_\mu$, we obtained the following equation of motion (EoM),
\begin{eqnarray}
&&R_{\mu\nu}-\frac{1}{2}\mathrm{g}_{\mu\nu}R= T_{\mu\nu},\label{egb2}\\
&& \nabla_{\mu}\left(\frac{\partial {{\cal L}_{ABG}(F)}}{\partial F}F^{\mu\nu}\right)=0\qquad \text{and} \qquad \nabla_{\mu}(* F^{\mu\nu})=0,
\label{fe}
\end{eqnarray}
where the matter energy-momentum tensor (EMT) is given by 
\begin{equation}
T_{\mu\nu}=2\left[\frac{\partial {{\cal L}_{ABG}(F)}}{\partial F}F_{\mu \sigma}F_{\nu}^{\sigma}-\mathrm{g}_{\mu\nu}{{\cal L}_{ABG}(F)}\right] +\frac{\rho \Omega_{\mu \sigma} \Omega^{\sigma}_{\phantom{\sigma} \nu}}{\sqrt{-\gamma}},
\end{equation} 
 {Here, we note that $\rho$ characterises the cloud of strings while $m$ signifies the mass of a single string (for details, see, e.g. \cite{lt2}).}
 The monopole's topological defects, like cosmic strings and domain walls, originated during the cooling phase of the early universe \cite{bar,kibb76} and play a significant role in investigating {BHs}.
%%%%%%%%%%%%%%%%%%%%%%%%%%%%%%%%%%%%%%%%%%%%%%%%%%
{ The non-vanishing component of the {EMT} is
\begin{eqnarray}
&&T^t_t=T^r_r=\frac{g^2(r^2-3g^2)}{(r^2+g^2)^3}-\frac{6M g^2}{(r^2+g^2)^{5/2}}+\frac{a}{r^2},\\
&&T^{\theta}_{\theta}= T^{\phi}_{\phi}=\frac{g^2(3g^4-8g^2r^2+r^4)}{(g^2+r^2)^4}+\frac{3g^2M (2g^4-g^2 r^2-3 r^4)}{(g^2+r^2)^{9/2}}{,}
\end{eqnarray}
where $a$ is a constant identified as the CS parameter. We consider the general static and the spherically symmetric line element to find the BH solution as 
\begin{equation}
ds^2=-f(r)dt^2+\frac{1}{f(r)}dr^2+r^2 \left(d\theta^2+\sin^2\theta d\phi^2\right),
\label{met1}
\end{equation}
where
\begin{equation}
     f(r)=1-\frac{2m(r)}{r}.
\end{equation}
The Einstein field equation is 
\begin{eqnarray}
m'(r)=\frac{g^2 r^2(r^2-3g^2)}{2(r^2+g^2)^3}- \frac{3M r^2 g^2}{(r^2+g^2)^{5/2}}+\frac{a}{2}.
\label{eom1}
\end{eqnarray}
Integrating Eq. (\ref{eom1}) from $r$ to $\infty$, we get,
\begin{equation}
- m(r)+\text{Constant}=  \int_r^{\infty} dr\left[ \frac{g^2 r^2(r^2-3g^2)}{2(r^2+g^2)^3}+ \frac{3M r^2 g^2}{(r^2+g^2)^{5/2}}+\frac{a}{2}\right],
\end{equation}
and 
\begin{equation}
    \text{Constant}={\lim}_{r \to \infty} \left[m(r)\right]= M.
\end{equation}
Finally, $m(r)$ is given by
\begin{equation}
    m(r)=\frac{Mr^3}{(r^2+g^2)^{3/2}}-\frac{g^2r^3}{2(r^2+g^2)^2}+\frac{a}{2}r{,}
\end{equation}
and the solution is given by
\begin{eqnarray}
f(r)= 1-\frac{2M r^2}{(r^2+g^2)^{3/2}}+\frac{g^2r^2}{(r^2+g^2)^2}-a.
\label{bhs}
\end{eqnarray}
}

%%%%%%%%%%%%%%%%%%%%%%%%%%%%%%%%%%%%%%%%%%%%%%%%%%%

 The solution (\ref{bhs}) is characterised by BH mass  ($M$),  MM  charge ($g$), and CS parameter ($a$). This exact solution elucidates a new BH solution in the presence of NLED (ABG source) and CS.   This solution is a generalised version of the Letelier solution \cite{ay1,xu19} in the absence of MM charge ($g$), and it interpolates with the  ABG BH solution in the absence of  CS ($a$) \cite{Cai:2021ele} (when $\alpha=3, \beta=4$).    this solution (\ref{bhs}) resemble  with the Schwarzschild  BH solution when $a=g=0$.  
 
{ Now, let us discuss the horizon structure of the obtained BH solution (\ref{bhs}). The horizon radii are zeros of $g^{
rr} = 0$ of $f(r_h)= 0$ , which implies that}
\begin{equation}
 1-\frac{2M  r_h^2}{(r_h^2+g^2)^{3/2}}+ \frac{g^2r_h^2}{(r_h^2+g^2)^2}-a =0,
\label{hor}
\end{equation}
{  The  Eq. (\ref{hor}) cannot be solved analytically, and we need to solve it numerically, and the graphical results are illustrated in Fig. \ref{fig:1}.
\begin{figure*}[ht]
\begin{tabular}{c c c c}
\includegraphics[width=.5\linewidth]{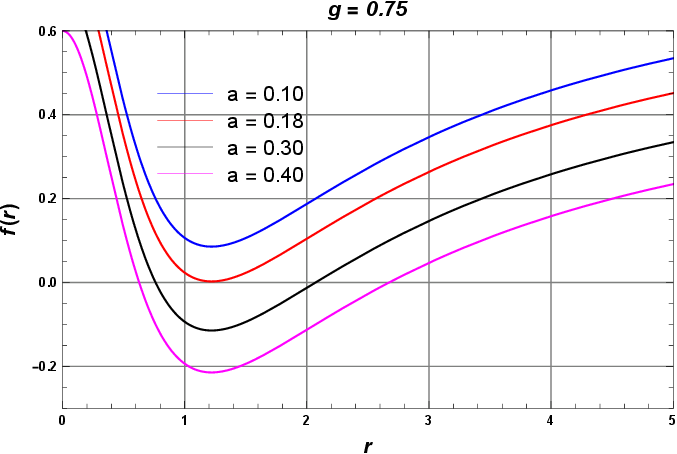}
\includegraphics[width=.5\linewidth]{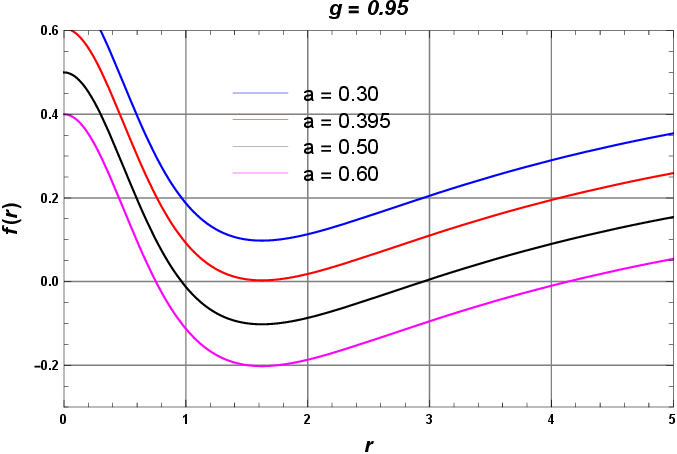}\\
\includegraphics[width=.5\linewidth]{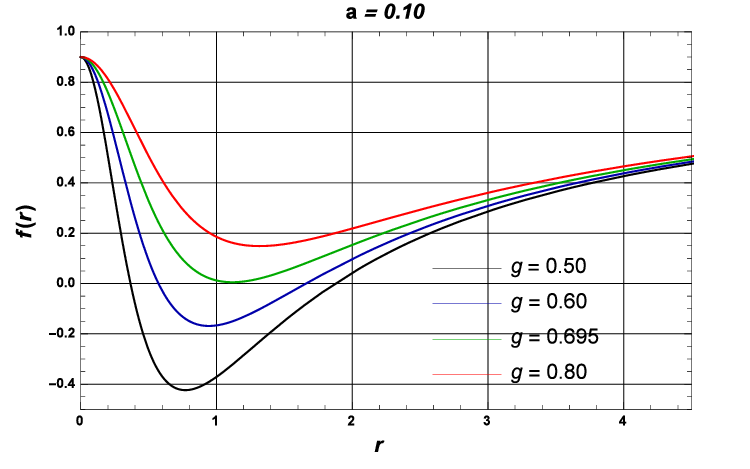}
\includegraphics[width=.5\linewidth]{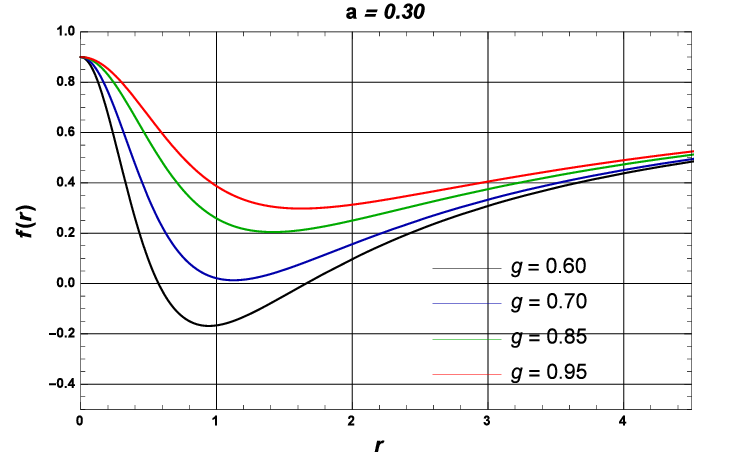}\\
\end{tabular}
\caption{The plot of the metric function $f(r)$ versus  $r$. Upper Panel: For various values of the CS parameter but fixed MM charge ($g=0.75$ and $g=0.95$). Lower Panel: For various values of MM charge but fixed  CS parameter ($a=0.1$   and  $a=0.3$) in a unit of $M $, where $M=1$.}
\label{fig:1}
\end{figure*} The metric function versus radial coordinate is plotted for various values of the CS parameter ($a$) and MM charge  ($g$). Such that  Eq. (\ref{hor}) admits eight roots (six negative and two positive). The two positive roots $r_{\pm}$, with $r_-$ and $r_+$, represent the Cauchy and event horizons. By keeping the CS parameter (or MM charge) fixed, we find the value of the critical CS parameter ($a_c$) (or MM charge ($g_c$)). The Cauchy and event horizons coincide when $r_- =r_+$ corresponds to the extremal BH. The obtained BH solution has two horizon when $a>a_c$ ($g<g_c$) and no horizon when $a<a_c$ ($g<g_c$) (see the table \ref{tab:h1} and table \ref{tab:h2})}. Here, one can see that the    BH horizon decreases with the growing   CS parameter, $a$, and increases with the growing MM charge, $g$. The BH horizon coincides at the critical value of MM charge $(g_c=0.695)$ at $a=0.1$, and the BH horizon is known as the degenerate horizon. For the value of $g>0.85$, no BH solutions exist. The critical MM charge,  $g_c$, is an increasing function of the CS parameter, $a$. We have noticed the effect of MM charge and CS parameter are opposite (see Fig. \ref{fig:1}).
\begin{center}
	\begin{table}[h]
		\begin{center}
			\begin{tabular}{l l r l| r l r l r}
				\hline
				\hline
				\multicolumn{1}{c}{ }&\multicolumn{1}{c}{ $g=0.75$  }&\multicolumn{1}{c}{}&\multicolumn{1}{c|}{ \,\,\,\,\,\, }&\multicolumn{1}{c}{ }&\multicolumn{1}{c}{}&\multicolumn{1}{c}{ $g=0.95$ }&\multicolumn{1}{c}{}\,\,\,\,\,\,\\
				\hline
				\multicolumn{1}{c}{ \it{$a$}} & \multicolumn{1}{c}{ $r_-$ } & \multicolumn{1}{c}{ $r_+$ }& \multicolumn{1}{c|}{$\delta$}&\multicolumn{1}{c}{$a$}& \multicolumn{1}{c}{$r_-$} &\multicolumn{1}{c}{$r_+$} &\multicolumn{1}{c}{$\delta$}   \\
				\hline

	\,\,\, 0.183\,\,& \,\,1.21\,\, &\,\,  1.21\,\,& \,\,0\,\,&0.395&\,\, 1.615\,\,&\,\,1.615\,\,&\,\,0\,\,
				\\
	\,\, 0.30\,\, & \,\,0.761\,\, &\,\, 2.07\,\,& \,\,1.309\,\,&0.50&\,\, 1.749\,\,&\,\,1.509\,\,&\,\,0.240\,\,
				\\
	\,\,\, 0.40\,\, &  \,\,0.627\,\,  &\,\,2.67\,\,&\,\,0.943\,&0.50\,& \,\, 2.945\,\,&\,\,0.964\,\,&\,\,0.981\,\,
				\\
				\hline
				\hline
			\end{tabular}
		\end{center}
		\caption{ The numerical values of inner  horizon ($r_-$), outer horizon ($r_+$) and $\delta=r_+-r_-$ for various CS parameter $a$ for $g=0.75$ and $g=0.95$ with fixed   mass $(M=1)$.}
				\label{tab:h1}
	\end{table}
\end{center}

\begin{center}
	\begin{table}[h]
		\begin{center}
			\begin{tabular}{l l r l| r l r l r}
				\hline
				\hline
				\multicolumn{1}{c}{ }&\multicolumn{1}{c}{ $a=0.10$  }&\multicolumn{1}{c}{}&\multicolumn{1}{c|}{ \,\,\,\,\,\, }&\multicolumn{1}{c}{ }&\multicolumn{1}{c}{}&\multicolumn{1}{c}{ $a=0.20$ }&\multicolumn{1}{c}{}\,\,\,\,\,\,\\
				\hline
				\multicolumn{1}{c}{ \it{$g$}} & \multicolumn{1}{c}{ $r_-$ } & \multicolumn{1}{c}{ $r_+$ }& \multicolumn{1}{c|}{$\delta$}&\multicolumn{1}{c}{$g$}& \multicolumn{1}{c}{$r_-$} &\multicolumn{1}{c}{$r_+$} &\multicolumn{1}{c}{$\delta$}   \\
				\hline

	\,\,\, 0.50\,\,& \,\,0.366\,\, &\,\,  1.875\,\,& \,\,1.509\,\,&0.60&\,\, 0.432\,\,&\,\,2.425\,\,&\,\,0.590\,\,
				\\
	\,\, 0.60\,\, & \,\,0.572\,\, &\,\, 1.659\,\,& \,\,1.087\,\,&0.70&\,\, 0.626\,\,&\,\,1.216\,\,&\,\,0.388\,\,
				\\
	\,\,\, 0.695\,\, &  \,\,1.11\,\,  &\,\,1.11\,\,&\,\,0\,&0.85\,& \,\, 1.407\,\,&\,\,1.407\,\,&\,\,0\,\,
				\\

				\hline
				\hline
			\end{tabular}
		\end{center}
		\caption{Inner  horizon ($r_-$), outer horizon ($r_+$) and $\delta=r_+-r_-$ for various   MM charge $g$ for $a=0.10$ and $a=0.20$ with fixed  mass $(M=1)$.}
				\label{tab:h2}
	\end{table}
\end{center}

\section{Thermodynamics}\label{sec3}
In this section, we discuss the thermodynamic properties of the 
ABG BH with CS. To serve the purpose, let us first 
calculate the gravitation mass of the BH with the help of the horizon condition ($f(r)|_{r=r_+}=0$)  as 
\begin{eqnarray}
M_+  =\frac{1}{\sqrt{r_+^2+g^2}}\left(\frac{3g^2}{2}-a g^2+\frac{g^4}{2r_+^2}-\frac{ag^4}{2r_+^2}-\frac{ar_+^2}{2}+\frac{r_+^2}{2}\right).
\label{eq:mass}
\end{eqnarray}
The Hawking temperature is calculated as 
\begin{eqnarray}
T_+= \frac{f'(r)}{4\pi}=\frac{1}{4\pi r_+(r_+^2+g^2)^3}\left(r_+^6-2g^6-3g^4r_+^2-g^2r_+^4+a(2g^6+3g^4r_+^2-r_+^6)\right).
\label{eq:temp}
\end{eqnarray}
 \begin{figure*}[ht]
\begin{tabular}{c c c c}
\includegraphics[width=.5\linewidth]{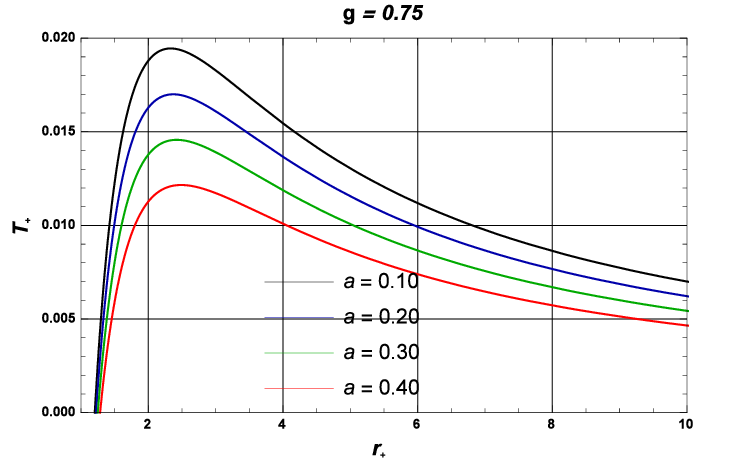}
\includegraphics[width=.5\linewidth]{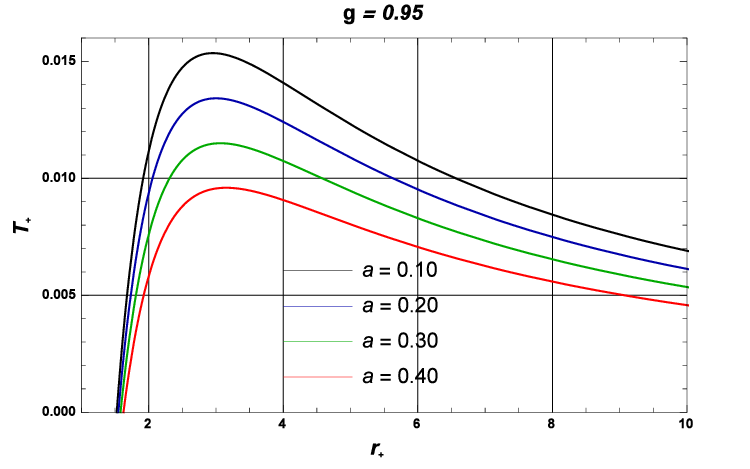}\\
\includegraphics[width=.5\linewidth]{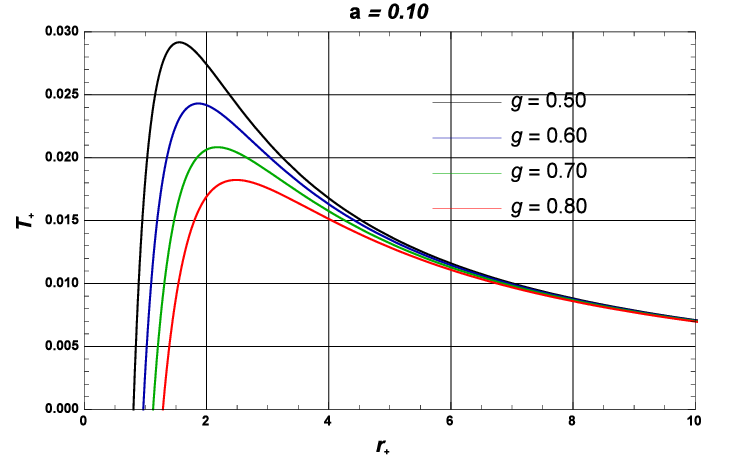}
\includegraphics[width=.5\linewidth]{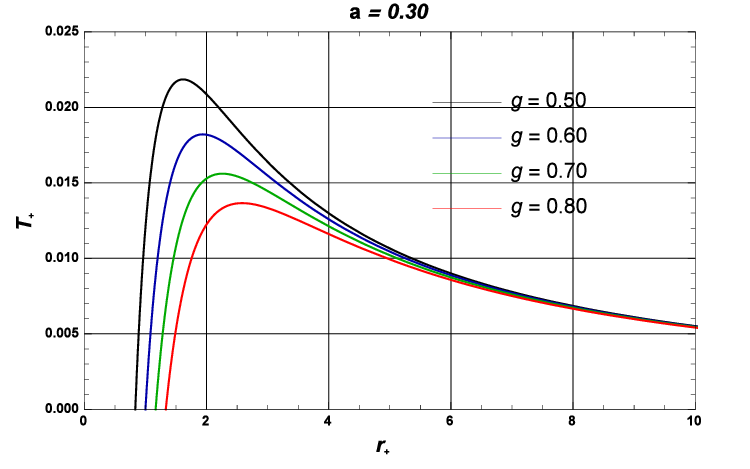}\\
\end{tabular}
\caption{The plot of temperature $T_+$ vs horizon radius $r_+$. Upper Panel: For various CS parameter values with a fixed MM charge value ($g=0.75$  and  $g=0.95$). Lower Panel: For {various} values of MM charge with a fixed value of CS parameter ($a=0.1$  and  $a=0.2$) in the unit of $M$, where $M=1$.}
\label{fig:t}
\end{figure*}
The temperature of the obtained BH solution  interpolates with the  Letelier BH  in the limit of $g= 0$ \cite{uma}
\begin{eqnarray}
&&M_+  =\frac{r_+}{2}(1-a),\\
&&T_+= \frac{1-a}{4\pi r_+}.
\end{eqnarray}
These quantities identify the    ABG BH when the CS parameter is switched off \cite{Singh:2021nvm,xu19}, i.e.,
\begin{eqnarray}
&&M_+  =\frac{1}{\sqrt{r_+^2+g^2}}\left(\frac{3g^2}{2}+\frac{g^2}{2r_+^2}+\frac{r_+^2}{2}\right),\\
&&T_+= \frac{1}{4\pi r_+(r_+^2+g^2)^3}\left(r_+^6-2g^6-3g^4r_+^2-g^2r_+^4\right).
    \label{hc3}
\end{eqnarray}
However, these quantities coincide with Schwarzschild BH when both the parameters $g=a= 0$.

The entropy of the obtained BH solution can be calculated as 
this must satisfy the first law of thermodynamics
\begin{equation}
dM_+ =T_+dS_+ + \Phi dg.
\end{equation}
At constant charge, the entropy is calculated as
\begin{eqnarray}
S_+=\int \frac{1}{T_+}\frac{dM_+ }{dr_+} dr_+=  \pi r_+ {(1-\frac{2g^2}{r_+^2})\sqrt{r_+^2+g^2}} - {3\pi g^2} \ln\left[\sqrt{r_+^2+g^2}-r_+\right].
\label{ent}
\end{eqnarray}
where 
the terms in bracket of Eq. (\ref{ent}) {are} due to NLED, which modifies the area law. Here, it is evident that the entropy does not follow the area law.  We can also derive the temperature according to the entropy using the first law of thermodynamics
\begin{equation}
    T_+= \frac{\partial M}{\partial S}=\frac{1}{4\pi r_+^4(r_+^2+g^2)^{3/2}}\left(r_+^6-2g^6-3g^4r_+^2-g^2r_+^4+a(2g^6+3g^4r_+^2-r_+^6)\right).
    \label{tempm}
\end{equation}
We have calculated the temperature in Eq. (\ref{eq:temp}) and Eq. (\ref{tempm}) of the obtained BH by different methods. The deviation relies on the general structure of the EMT of matter fields. When the BH mass parameter $M$ is included in the EMT, the conventional form of the first law gets modified with an extra factor \cite{ma14}
\begin{equation}
 \mathcal{C}(M_+,g,r_+)\,dM_+=T_+ \,dS,\label{cor}
\end{equation}
where $T_+$
 is the Hawking temperature and $\mathcal{C}(M_+,g,r_+)$ is
\begin{equation}
 \mathcal{C}(M_+,g,r_+)=1+4\pi \int_{r_+}^{\infty}r_+^2\frac{\partial T^t_t}{\partial M_+} dr_+= \frac{r_+^3}{(r_+^2+g^2)^{3/2}}.
\end{equation}
We recover the conventional form of the first law of BH thermodynamics when the factor ($\mathcal{C}(M_+,r_+)$=1) because the energy-momentum tensor does not depend upon mass.  Because any  BH
has temperature, it can be seen as a thermodynamic system. Thus, the conventional thermodynamic laws must be satisfied. We have two choices to connect Eq. (\ref{cor}) with the first law of thermodynamics. We know that $\delta E =T_+ \delta S_+$  then the $E\to M$ and the  entropy becomes 
\begin{equation}
 \delta S_+= \frac{\mathcal{C}(M_+,g,r_+)}{T_+}\delta M_+,
\end{equation}
and the the temperature in Eq. (\ref{eq:temp}) and Eq. (\ref{tempm}) of the obtained BH solution are same.

Following  this modified first law of thermodynamics, the entropy reads
\begin{equation}
S_+=\pi r_+^2.
\label{modent}
\end{equation}
Now, this entropy agrees with the area law and matches precisely with the entropy of  BHs.

{
\section{Local and Global Stability}
The heat capacity ($C_+$) and  Gibbs free energy $(G_+)$ study the system's local and global stability. The system is stable when $C_+>0$ ($G_+<0$) and unstable when $C_+<0$ ($G_+>0$). The following relation calculates the heat capacity  ($C_+$) of the BH
\begin{equation}
C_+=\frac{dM_+}{dT_+}=\frac{dM_+}{dr_+}\frac{dr_+}{dT_+}{.}
\label{eq:hc}
\end{equation}
Substituting the value of the ($M_+$) from Eq. (\ref{eq:mass}) and ($T_+$)  from Eq. (\ref{eq:temp}) in Eq. (\ref{eq:hc}), we get
\begin{equation}
C_+=-\frac{2\pi (g^2+r_+^2)^{5/2}[(1-a)(2g^6+3g^4r_+^2-r_+^6)+g^2r_+^4]}{r_+[(1-a)(2g^8+11g^6r_+^2-r_+^8)+3(4-5a)g^4r_+^4+(8-5a)g^2r_+^6]}{.}
\end{equation}
 \begin{figure*}[ht]
\begin{tabular}{c c c c}
\includegraphics[width=.5\linewidth]{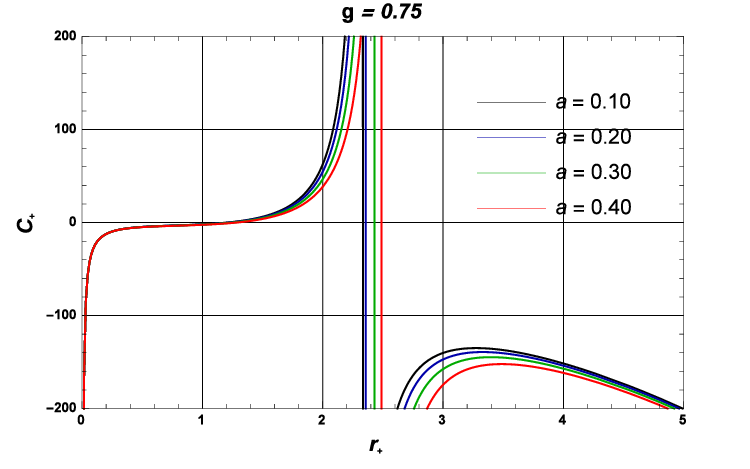}
\includegraphics[width=.5\linewidth]{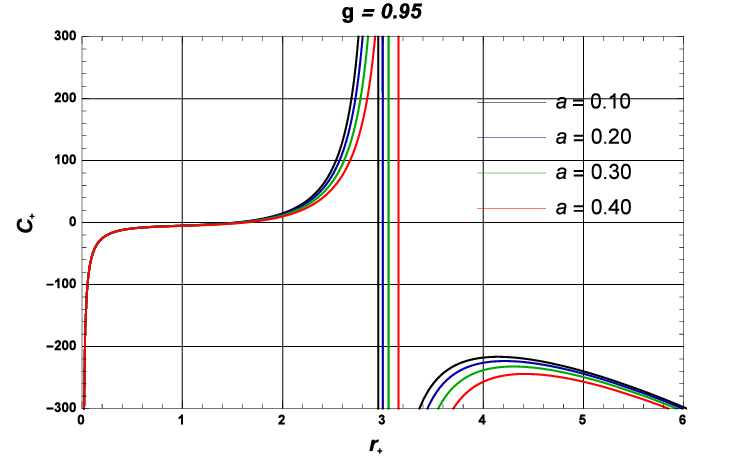}\\
\includegraphics[width=.5\linewidth]{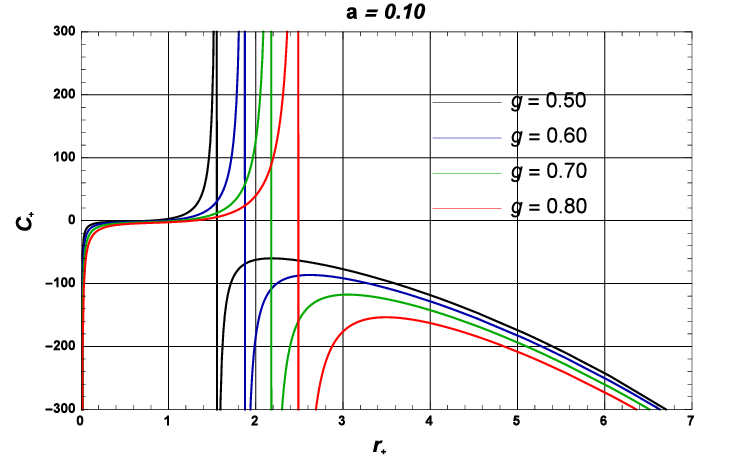}
\includegraphics[width=.5\linewidth]{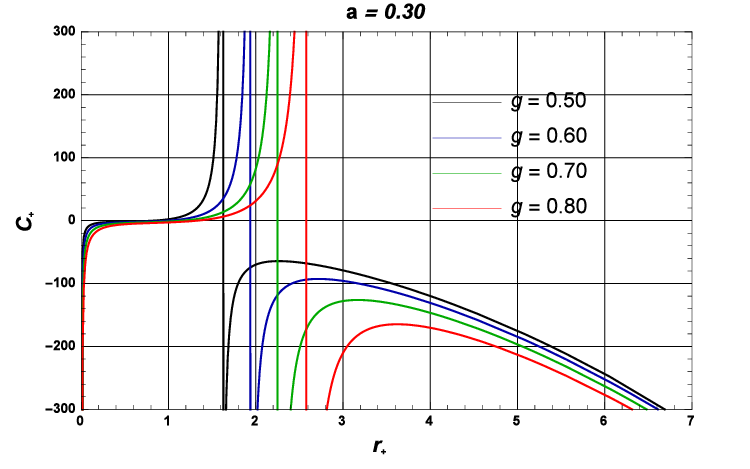}\\
\end{tabular}
\caption{The plot of heat capacity $C_+$ vs horizon radius $r_+$ Upper Panel: For various values of CS parameter with a fixed value of MM charge ($g=0.75$) and ($g=0.95$). Lower Panel: For {various} values of MM charge with a fixed value of CS parameter ($a=0.1$) and ($a=0.3$).}
\label{fig:c}
\end{figure*}
The heat capacity of the obtained BH solution is plotted in Fig. \ref{fig:c}. We observe two
kinds of behaviour: first is the positive heat capacity $r<r_c$, suggesting the thermodynamic stability of a BH, and
the other is negative heat capacity $r>r_c$, indicating instability of BH. The heat capacity is discontinuous at
$r=r_c$, which means that second-order phase transition occurs \cite{dharm, page}. It may be noted
that the critical radius $r_c$ changes drastically in the presence of the MM charge, and the CS parameter
increases. The critical radius rises with the MM charge and CS parameter increase.

 The following relation calculates the Gibbs free energy  ($G_+$) of the BH:
\begin{equation}
G_+=M_+-T_+S_+.
\label{eq:g}
\end{equation}
Substituting the value of the ($M_+$) from Eq. (\ref{eq:mass}) and ($T_+$)  from Eq. (\ref{eq:temp}) in Eq. (\ref{eq:g}), we get
\begin{equation}
G_+=\frac{(1-a)(r_+^2+g^2)+r_+^2g^2(3-2a)}{2r_+^2\sqrt{g^2+r_+^2}}+\frac{r_+[(1-a)(2g^6+2g^4r_+^2-r_+^6)+g^2r_+^4]}{4(g^2+r_+^2)^3}{.}
\label{eq:hc1}
\end{equation}
 \begin{figure*}[ht]
\begin{tabular}{c c c c}
\includegraphics[width=.5\linewidth]{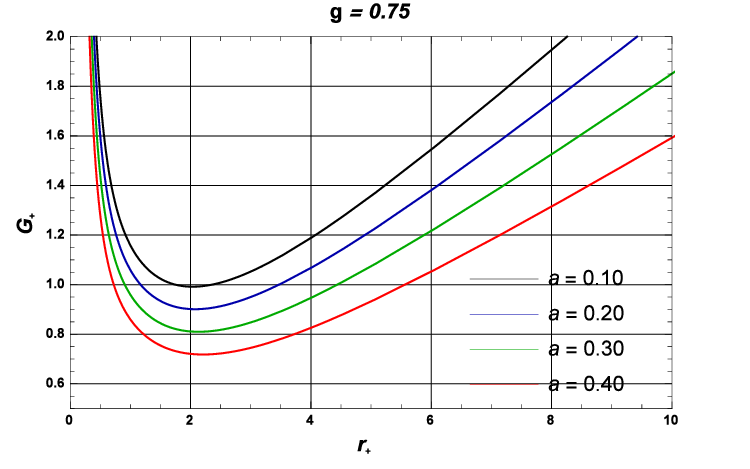}
\includegraphics[width=.5\linewidth]{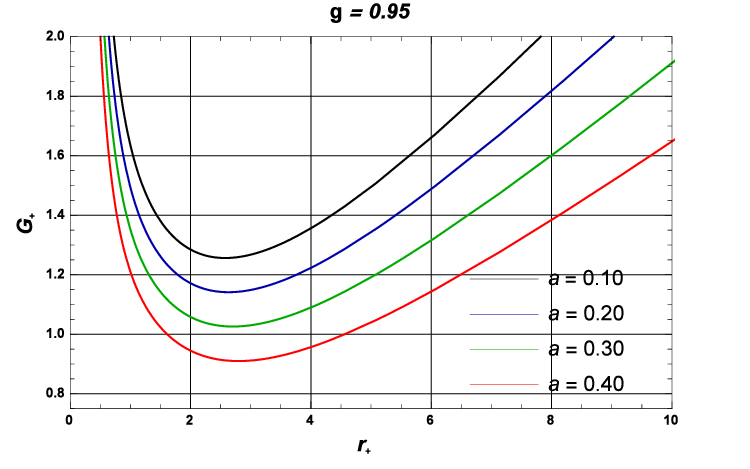}\\
\includegraphics[width=.5\linewidth]{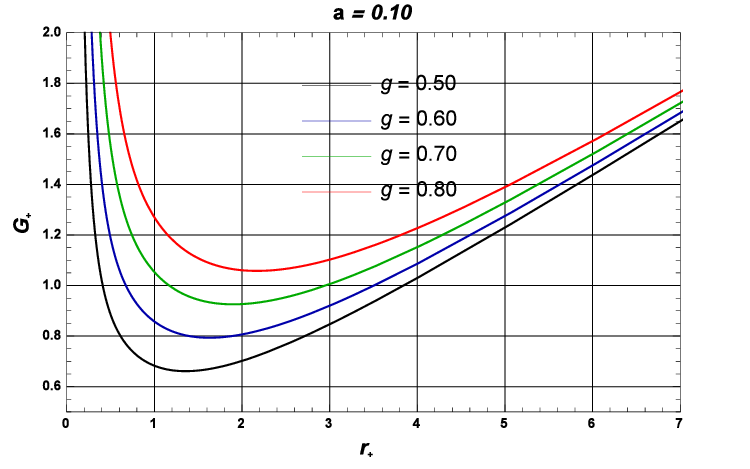}
\includegraphics[width=.5\linewidth]{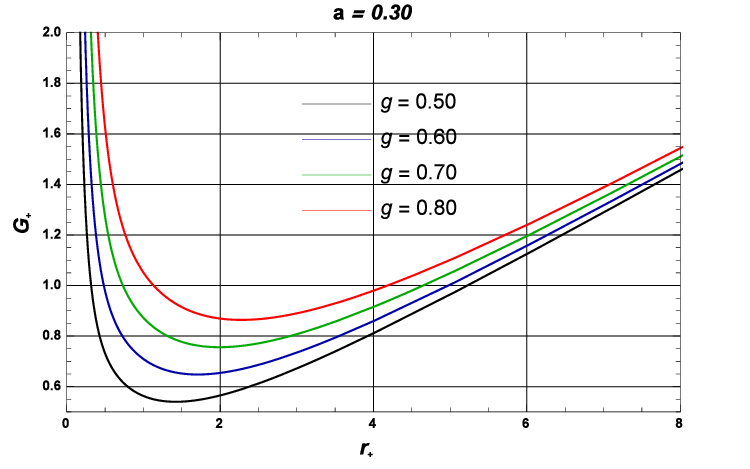}\\
\end{tabular}
\caption{The plot of heat capacity $C_+$ vs horizon radius $r_+$ Upper Panel: For various values of CS parameter with a fixed value of MM charge ($g=0.75$) and ($g=0.95$). Lower Panel: For {various} values of MM charge with a fixed value of CS parameter ($a=0.1$) and ($a=0.3$).}
\end{figure*}

Interestingly, the discontinuity of the heat
capacity occurs at $r=1.6$ for $g=0.60$ and $a=0.10$, at which point the Hawking temperature reaches the maximum value, and the Gibbs free energy reaches the minimum value. Hence, the phase transition occurs from the
lower to higher mass, corresponding to a black hole's positive to negative heat capacity.

The BH remnant is a well-merited entity in theoretical astrophysics that can be a source of dark energy \cite{jh87} and is one of the candidates to
resolve the information loss puzzle \cite{p92}. The double root $r=r_e$ of
$f(r)=0$ corresponds to the extremal BH with
a degenerate horizon. Hence $f'(r)=0$, and the critical radius is

\begin{equation}
r_e=\frac{g}{(1-a)}\sqrt{\left(\frac{1}{3}-\frac{2 2^{1/3} g^2}{A}\left[\frac{5}{3}+\frac{3a^2}{2}-3a\right]-\frac{A}{3 2^{1/3}}\right)}{,}
\label{eq:rcr}
\end{equation}
where
\begin{equation}
A= 3M^2\left(2g^{12}+27g^8M^4-54 g^6 M^6 +3\sqrt{3}\sqrt{-8g^{18}M^6-9g^{16}M^8-108 g^{14}M^{10}}\right)^{1/3}{,}
\end{equation}
and the corresponding mass is

\begin{equation}
M_e=\frac{1}{\sqrt{r_e^2+g^2}}\left(\frac{g^2(3-2a)}{2}+\frac{(g^4+r_e^4)(1-a)}{2r_e^2}\right){.}
\end{equation}
It can be seen from (\ref{eq:temp}) that the temperature decreases with increasing $r_+$  and vanishes when the two horizons coincide $T_+\to 0, C_+ \to 0$.
}
%-------------------------------------------------------------
\section{Shadow and Quasinormal Modes}\label{sec4}
In this section, we study the shadow and quasinormal modes of the 
BH solution coupled with ABG NLED and CS.
We begin by assuming that a photon is moving around the BH. The photon is restricted to move in the equatorial plane (i.e. $\theta=\pi/2$). The Hamiltonian that leads to the equation of motion \cite{Belhaj:2020okh,Belhaj:2022kek,Belhaj:2021tfc},
\begin{equation}
H=\frac{1}{2}\left[-\frac{p_t^2}{f(r)}+f(r)p_r^2+\frac{p_{\phi}^2}{r^2}\right].
\end{equation}
The canonically conjugate momenta corresponding to the line element (\ref{met1}) with    metric function (\ref{bhs}) are
\begin{eqnarray}
&&p_t=\left(1-\frac{2M r^2}{(r^2+g^2)^{3/2}}+\frac{g^2r^2}{(r^2+g^2)^2}-a\right){\dot t}={\cal E},\\&&p_r=\left[\left(1-\frac{2M r^2}{(r^2+g^2)^{3/2}}+\frac{g^2r^2}{(r^2+g^2)^2}-a\right)\right]^{-1}{\dot r},\\&&p_{\theta}=r^2{\dot\theta},\qquad \text{and}\qquad p_{\phi}=r^2\sin^2\theta {\dot \phi}=L.
\end{eqnarray}
Here, ${\cal E}$ and $L$  represent the energy and angular momentum of the photon, respectively. 

By exploiting the equations of motion and conserved quantities, the
radial null circular geodesics reads
\begin{equation}
    {\dot r}^2+V_{eff}(r)=0 \qquad \text{where} \qquad V_{eff}=f(r)\left(\frac{L^2}{r^2}+\frac{{\cal E}^2}{f(r)}\right).
\end{equation}
For null circular geodesics,  $V_{eff}$ must satisfy the following conditions:
\begin{equation}
V_{eff}=0, \qquad   \frac{\partial V_{eff}}{\partial r}=0,
\label{pot}
\end{equation}
which lead to
\begin{equation}
3 M r^4 (r^2+g^2)-\sqrt{r^2+g^2}(2g^2r^4+(1-a)(g^2+r^2)^3)=0.
\label{rp}
\end{equation}
The solution of this equation gives photon radius. However, this equation is not solvable analytically; therefore, this can be solved  
numerically. The numerical photon radius values for different MM charges and CS  parameter values are appended in table \ref{tr1}. 
\begin{table}[ht]
 \begin{center}
 \begin{tabular}{ |l | l   | l   | l   |  l |  l | l | l | l | l | }
\hline
            \hline
  \multicolumn{1}{|c}{ } &\multicolumn{1}{c}{}  &\multicolumn{1}{c}{}  &\multicolumn{1}{c }{ }&\multicolumn{1}{c }{ }&\multicolumn{1}{c }{ $r_p$}&\multicolumn{1}{c }{ }&\multicolumn{1}{c }{ }&\multicolumn{1}{c }{ } &\multicolumn{1}{c|}{}\\
            \hline
  \multicolumn{1}{|c|}{ $a$} &\multicolumn{1}{c|}{$g=0.1$}  &\multicolumn{1}{c|}{$g=0.2$}  &\multicolumn{1}{c|}{$g=0.3$} &\multicolumn{1}{c|}{$g=0.4$}&\multicolumn{1}{c|}{$g=0.5$}&\multicolumn{1}{c|}{$g=0.6$}&\multicolumn{1}{c|}{$g=0.7$}&\multicolumn{1}{c|}{$g=0.8$}&\multicolumn{1}{c|}{$g=0.9$}\\
            \hline
\,\,\,\,\,0.1 ~  &~3.319~  & ~3.736~ & ~4.273~ & ~4.988~& ~5.989~& ~7.489~& ~9.990~& ~14.991~& ~29.992~ \\            
\,\,\,\,\,0.2~  &~3.275~  & ~3.695~ & ~4.235~ & ~4.952~& ~5.956~& ~7.459~& ~9.963~& ~14.966~& ~29.970~ \\ 
\,\,\,\,\,0.3 ~  &~3.200~  & ~3.625~ & ~4.170~ & ~4.892~& ~5.900~& ~7.408~& ~9.916~& ~14.924~& ~29.932~ \\ 
\,\,\,\,\,0.4 ~  &~3.080~  & ~3.522~ & ~4.075~ & ~4.805~& ~5.821~& ~7.336~& ~9.851~& ~14.865~& ~29.879~ \\ 
\,\,\,\,\,0.5~  &~2.928~  & ~3.379~ & ~3.945~ & ~4.688~& ~5.715~& ~7.241~& ~9.765~& ~14.788~& ~29.727~ \\ 
\,\,\,\,\,0.6 ~  &~2.697~  & ~3.181~ & ~3.773~ & ~4.536~& ~5.580~& ~7.121~& ~9.658~& ~14.693~& ~29.811~ \\ 
\,\,\,\,\,0.7 ~  &~2.318~  & ~2.897~ & ~3.541~ & ~4.340~& ~5.411~& ~6.973~& ~9.529~& ~14.580~& ~29.628~ \\ 
\,\,\,\,\,0.8 ~  &~\ldots~  & ~2.386~ & ~3.211~ & ~4.084~& ~5.200~& ~6.794~& ~9.374~& ~14.446~& ~29.512~ \\ 
\,\,\,\,\,0.9 ~  &~\ldots~  & ~\ldots~ & ~2.582~ & ~3.729~& ~4.933~& ~6.577~& ~9.193~& ~14.293~& ~29.379~ \\ 
            \hline 
\hline
        \end{tabular}
        \caption{The numerical values of photon radius with  different values of $a$ and  $g$ in unit of $M$, where $M=1$.}
\label{tr1}
    \end{center}
\end{table}
From the list, we notice that the effects of the CS parameter, $a$, and MM charge, $g$, on photon radii are in contrast. The photon radius increases when the CS parameter increases, $a$, but decreases when the MM charge, $g$, increases. 
  
\subsection{Black Hole Shadow}
\begin{figure*}[h]
\begin{tabular}{c c c c}
\includegraphics[width=.5\linewidth]{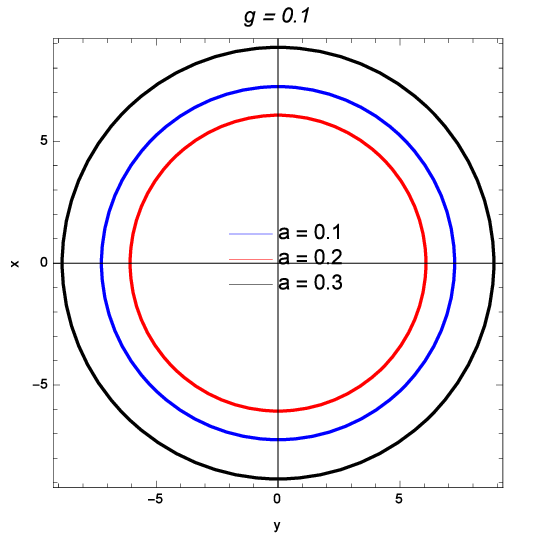}
\includegraphics[width=.5\linewidth]{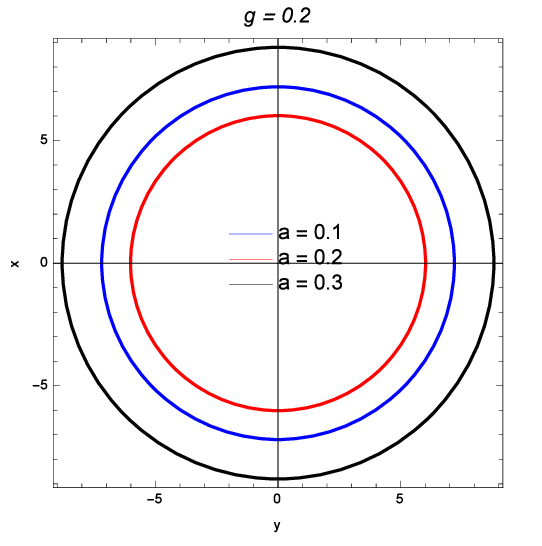}\\
\includegraphics[width=.5\linewidth]{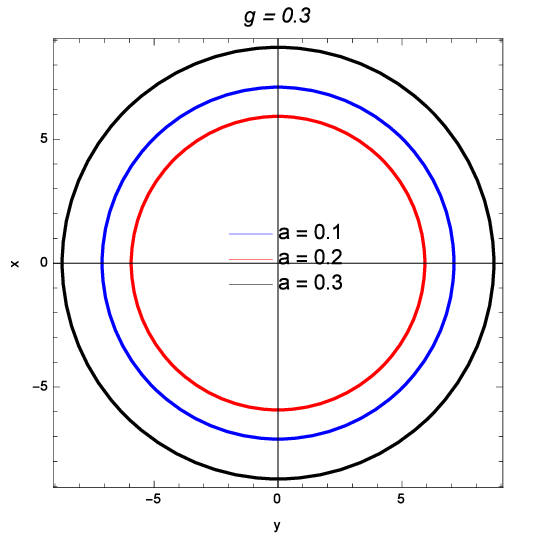}
\includegraphics[width=.5\linewidth]{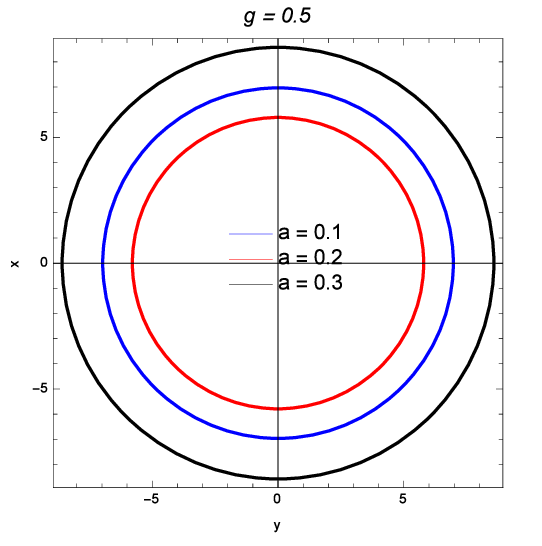}
\end{tabular}
\caption{ The  BH shadow   for different CS parameter  ($a$) and MM charge  ($g$).}
\label{fig2}
\end{figure*}

With the help of photon radius,  we can now compute the 
shadow of a  BH  (\ref{bhs}). The  BH shadow radius ($r_s$) depends on
photon sphere radius as  \cite{72}
\begin{equation}
r_s =\frac{r}{\sqrt{f(r)}}|_{r=r_p}.
\end{equation}
The numerical values of the shadow radius are given in table \ref{tr2} and plotted in Fig. \ref{fig2} for various  BH parameters. Here, it is evident that the shadow radius increases along with the increasing CS parameter, $a$, but decreases with the increasing MM charge, $g$.
\begin{table}[ht]
 \begin{center}
 \begin{tabular}{ |l | l   | l   | l   |  l |  l | l | l | l | l | }
\hline
            \hline
  \multicolumn{1}{|c}{ } &\multicolumn{1}{c}{}  &\multicolumn{1}{c}{}  &\multicolumn{1}{c }{ }&\multicolumn{1}{c }{ }&\multicolumn{1}{c }{ $r_s$}&\multicolumn{1}{c }{ }&\multicolumn{1}{c }{ }&\multicolumn{1}{c }{ } &\multicolumn{1}{c|}{}\\
            \hline
  \multicolumn{1}{|c|}{ $a$} &\multicolumn{1}{c|}{$g=0.1$}  &\multicolumn{1}{c|}{$g=0.2$}  &\multicolumn{1}{c|}{$g=0.3$} &\multicolumn{1}{c|}{$g=0.4$}&\multicolumn{1}{c|}{$g=0.5$}&\multicolumn{1}{c|}{$g=0.6$}&\multicolumn{1}{c|}{$g=0.7$}&\multicolumn{1}{c|}{$g=0.8$}&\multicolumn{1}{c|}{$g=0.9$}\\
            \hline
\,\,\,\,\,0.1 ~  &~6.068~  & ~6.015~ & ~5.924~ &  ~5.789~& ~5.600~& ~5.338~& ~4.945~& ~\ldots~&~\ldots~ \\            
 \,\,\,\,\,0.2~  &~7.244~  & ~7.191~ & ~7.100~ & ~6.967~& ~6.785~& ~6.540~& ~6.203~& ~5.680~& ~\ldots~ \\ 
\,\,\,\,\,0.3 ~  &~8.854~  & ~8.801~ & ~8.710~ & ~8.578~& ~8.400~& ~8.166~& ~7.860~& ~7.445~& ~6.794~ \\ 
\,\,\,\,\,0.4 ~  &~11.162~  & ~11.108~ & ~11.016~ & ~10.884~& ~10.707~& ~10.480~& ~10.192~& ~9.825~& ~9.341~ \\ 
 \,\,\,\,\,0.5~  &~14.678~  & ~14.623~ & ~14.529~ & ~14.325~& ~14.218~& ~13.993~& ~13.714~& ~13.371~& ~12.946~ \\ 
 \,\,\,\,\,0.6 ~  &~20.520~  & ~20.462~ & ~20.365~ & ~20.227~& ~20.045~& ~19.818~& ~19.540~& ~19.206~& ~18.806~ \\ 
\,\,\,\,\,0.7 ~  &~31.602~  & ~31.540~ & ~31.436~ & ~31.289~& ~31.098~& ~30.860~& ~30.573~& ~30.233~& ~29.835~ \\ 
\,\,\,\,\,0.8 ~  &~58.071~  & ~58.001~ & ~57.884~ & ~57.720~& ~57.506~& ~57.243~& ~56.928~& ~56.559~& ~56.133~ \\ 
\,\,\,\,\,0.9 ~  &~164.28~  & ~164.19~ & ~164.04~ & ~163.83~& ~163.56~& ~163.22~& ~162.82~& ~162.36~& ~161.83~ \\ 
            \hline 
\hline
        \end{tabular}
        \caption{The numerical values of shadow radius for different values of   CS parameters ($a$) and MM charge ($g$) in a unit of $M$, where $M=1$.}
\label{tr2}
    \end{center}
\end{table}

%----------------------------------------------------------------------------------------------------------
\subsection{QNMs of  Black Hole Solution}
In this subsection, we utilise the scalar QNMs of the BH   to investigate the dynamical stability of the solution, and this is characterised by the real and imaginary parts of the complex QNM frequencies (QNFs),   $\omega=\omega_R+i \omega_I$. The condition $\omega>0$ confirms that the BH  is unstable, and condition $\omega<0$ indicates that the BH is stable. We consider scalar field perturbations of the BH solution.
The QNMs and  QNFs can be computed from the solution of the following scalar field equation for the BH background:
%------------------------------------------------------------n
\begin{equation}
\frac{1}{\sqrt{-\mathrm{g}}}\partial_{\mu}\left(\sqrt{-\mathrm{g}} \mathrm{g}^{\mu\nu}\partial_{\nu}\right)\psi=0.
\label{scalar1}
\end{equation}
%------------------------------------------------------------
This equation can be solved by separating the variables as, 
%------------------------------------------------------------
\begin{equation}
\psi=\frac{1}{r}\sum_{lm}e^{i\omega t} u_{lm}(r)Y^m_{l}(\theta,\phi),
\label{scalar2}
\end{equation}
%-----------------------------------------------------------
 where $Y^m_{l}$  are  spherical harmonics. For the tortoise 
 coordinate $dr^{*}=dr/f(r)$, the radial part of the solution takes the 
 Schr\"odinger-like form 
%--------------------------------------------------------------------
\begin{equation}
\left(\frac{d^2}{dr^{*^2}}+\omega^2-V_0(r^{*})\right) u(r)=0,
\end{equation}
%----------------------------------------------------------
where, $V_0(r^{*})=f(r)\left(\frac{f'(r)}{f(r)}+\frac{l(l+1)}{r^2}
\right)$.
 We use the WKB method to solve the QNFs in large $l$ limit 
\cite{will,iyer,konoplya1,wkb1} as 
\begin{equation}
\omega=l\Omega-i\left(n+\frac{1}{2}\right)|\Lambda|,    
\end{equation}
 with    
\begin{eqnarray}
 \Omega=\frac{\sqrt{f(r_p)}}{r_p}=\frac{1}{L_p}\qquad \text{and}\qquad \Lambda=\frac{\sqrt{2f(r_p)-r^2_pf''(r_p)}}{\sqrt{2} L_p}.
\end{eqnarray}

\begin{figure*}[ht]
\begin{tabular}{c c c c}
\includegraphics[width=.5\linewidth]{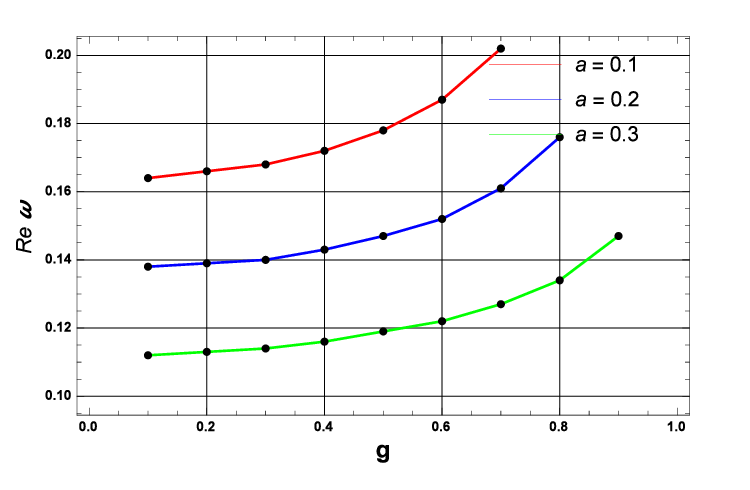}
\includegraphics [width=.5\linewidth]{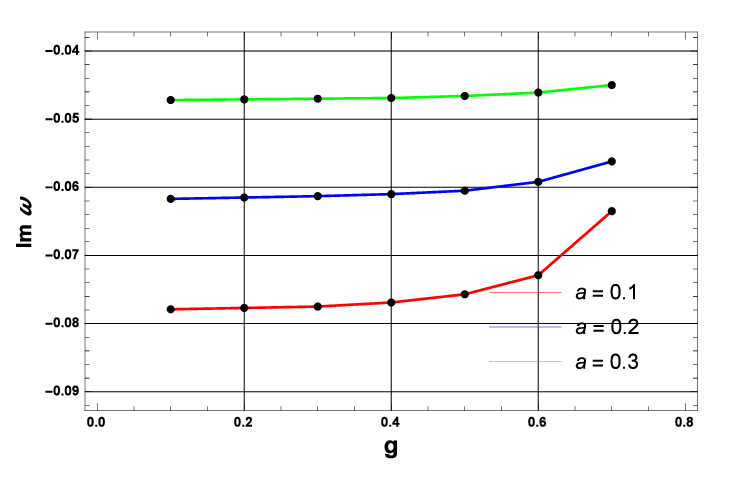}
\end{tabular}
\caption{Left panel: The real part of QNFs for various values of  MM charge  $(g)$  with fixed    $M$.  Right panel: The imaginary part of QNFs for various values of  MM charge  $(g)$ in a unit of $M$, where  $M=1$.}
\label{sh}
\end{figure*}

The numerical values of  QNFs are tabulated in tables \ref{tr14} and \ref{tr15} and plotted in Fig. \ref{sh} for the different BH parameters.
\begin{table}[ht]
 \begin{center}
 \begin{tabular}{ |l | l   | l   | l   |  l |  l | l | l | l | l | }
\hline
            \hline
  \multicolumn{1}{|c}{ } &\multicolumn{1}{c}{}  &\multicolumn{1}{c}{}  &\multicolumn{1}{c }{ }&\multicolumn{1}{c }{ }&\multicolumn{1}{c }{ $\omega_R$}&\multicolumn{1}{c }{ }&\multicolumn{1}{c }{ }&\multicolumn{1}{c }{ } &\multicolumn{1}{c|}{}\\
            \hline
  \multicolumn{1}{|c|}{ $a$} &\multicolumn{1}{c|}{$g=0.1$}  &\multicolumn{1}{c|}{$g=0.2$}  &\multicolumn{1}{c|}{$g=0.3$} &\multicolumn{1}{c|}{$g=0.4$}&\multicolumn{1}{c|}{$g=0.5$}&\multicolumn{1}{c|}{$g=0.6$}&\multicolumn{1}{c|}{$g=0.7$}&\multicolumn{1}{c|}{$g=0.8$}&\multicolumn{1}{c|}{$g=0.9$}\\
            \hline
\,\,\,\,\,0.1 ~  &~0.164~  & ~0.166~ & ~0.168~ &  ~0.172~& ~0.178~& ~0.187~& ~0.202~& ~\ldots~&~\ldots~ \\            
 \,\,\,\,\,0.2~  &~0.138~  & ~0.139~ & ~0.140~ & ~0.143~& ~0.147~& ~0.152~& ~0.161~& ~0.176~& ~\ldots~ \\ 
\,\,\,\,\,0.3 ~  &~0.112~  & ~0.113~ & ~0.114~ & ~0.116~& ~0.119~& ~0.122~& ~0.127~& ~0.134~& ~0.147~ \\ 
\,\,\,\,\,0.4 ~  &~0.0895~  & ~0.0900~ & ~0.0907~ & ~0.0918~& ~0.0933~& ~0.0954~& ~0.0981~& ~0.101~& ~0.107~ \\ 
 \,\,\,\,\,0.5~  &~0.0681~  & ~0.0683~ & ~0.0688~ & ~0.0694~& ~0.0703~& ~0.0714~& ~0.0729~& ~0.0747~& ~0.0722~ \\ 
 \,\,\,\,\,0.6 ~  &~0.0487~  & ~0.488~ & ~0.0491~ & ~0.0494~& ~0.0498~& ~0.0504~& ~0.0511~& ~0.0520~& ~0.0531~ \\ 
\,\,\,\,\,0.7 ~  &~0.0316~  & ~0.0317~ & ~0.0318~ & ~0.0319~& ~0.0321~& ~0.0324~& ~0.0327~& ~0.0330~& ~0.0335~ \\ 
\,\,\,\,\,0.8 ~  &~0.0172~  & ~0.0172~ & ~0.0172~ & ~0.0173~& ~0.0173~& ~0.0174~& ~0.0175~& ~0.0176~& ~0.0178~ \\ 
\,\,\,\,\,0.9 ~  &~0.006~  & ~0.006~ & ~0.006~ & ~0.006~& ~0.006~& ~0.006~& ~0.006~& ~0.006~& ~0.006~ \\ 
            \hline 
\hline
        \end{tabular}
        \caption{The numerical values of the real part of  QNMs for various    CS parameters, $a$ and MM charge, $g$ in a unit of $M$. Here,  $M=1$, $l=1$ and $n=0$.}
\label{tr14}
    \end{center}
\end{table}

\begin{table}[ht]
 \begin{center}
 \begin{tabular}{ |l | l   | l   | l   |  l |  l | l | l | l | l | }
\hline
            \hline
  \multicolumn{1}{|c}{ } &\multicolumn{1}{c}{}  &\multicolumn{1}{c}{}  &\multicolumn{1}{c }{ }&\multicolumn{1}{c }{ }&\multicolumn{1}{c }{ 
  $-\omega_I$}&\multicolumn{1}{c }{ }&\multicolumn{1}{c }{ }&\multicolumn{1}{c }{ } &\multicolumn{1}{c|}{}\\
            \hline
  \multicolumn{1}{|c|}{ $a$} &\multicolumn{1}{c|}{$g=0.1$}  &\multicolumn{1}{c|}{$g=0.2$}  &\multicolumn{1}{c|}{$g=0.3$} &\multicolumn{1}{c|}{$g=0.4$}&\multicolumn{1}{c|}{$g=0.5$}&\multicolumn{1}{c|}{$g=0.6$}&\multicolumn{1}{c|}{$g=0.7$}&\multicolumn{1}{c|}{$g=0.8$}&\multicolumn{1}{c|}{$g=0.9$}\\
            \hline
\,\,0.1 ~  &~  0.077~   & ~ 0.0777~ & ~ 0.0775~ &  ~ 0.0769~& ~ 0.0757~& ~ 0.0729~& ~ 0.0635~& ~ \ldots~&~\ldots~ \\            
 \,\,0.2~  &~ 0.061~  & ~0.0615~ & ~ 0.0613~ & ~ 0.0610~& ~ 0.0605~& ~ 0.0592~& ~ 0.0562~& ~ 0.0439~& ~\ldots~ \\ 
\,\,0.3 ~  &~ 0.047~  & ~ 0.0471~ & ~ 0.0470~ & ~ 0.0469~& ~ 0.0466~& ~ 0.0461~& ~ 0.0450~& ~ 0.0423~& ~ 0.0300~ \\ 
\,\,0.4 ~  &~ 0.034~  & ~ 0.0346~ & ~ 0.0346~ & ~ 0.0345~& ~ 0.0344~& ~ 0.0342~& ~ 0.0338~& ~ 0.0331~& ~ 0.0312~ \\ 
 \,\,0.5~  &~ 0.024~  & ~ 0.0240~ & ~ 0.0240~ & ~ 0.0240~& ~ 0.0240~& ~ 0.0239~& ~ 0.0238~& ~ 0.0236~& ~ 0.0232~ \\ 
 \,\,0.6 ~  &~ 0.015~  & ~ 0.153~ & ~ 0.0154~ & ~ 0.0154~& ~ 0.0153~& ~0.0153~& ~ 0.0153~& ~ 0.0153~& ~ 0.0152~ \\ 
\,\,0.7 ~  &~ 0.008~  & ~ 0.0086~ & ~0.0086~ & ~0.0086~& ~ 0.0086~& ~0.0086~& ~0.0086~& ~0.0086~& ~ 0.0086~ \\ 
\,\,0.8 ~  &~ 0.003~  & ~ 0.0038~ & ~ 0.0038~ & ~ 0.0038~& ~ 0.0038~& ~ 0.0038~& ~ 0.0038~& ~ 0.0038~& ~ 0.0038~ \\ 
\,\,0.9 ~  &~ 0.0009~  & ~ 0.0009~ & ~ 0.0009~ & ~ 0.0009~& ~ 0.0009~& ~ 0.0009~& ~ 0.0009~& ~ 0.0009~& ~ 0.0009~ \\ 
            \hline 
\hline
        \end{tabular}
        \caption{The numerical values of the imaginary part of QNMs for different  CS parameters, $a$ and MM charge, $g$ in a unit of $M$. Here,   $M=1$,   $l=1$ and $n=0$.}
\label{tr15}
    \end{center}
\end{table}

The behaviour of QNMs and QNFs under the influence of parameters $a$ and $g$ are depicted in Fig. \ref{sh}. Here, the fundamental part of the QNMs decreases with $a$ and increases with the $g$ while the imaginary part increases (almost constant) with $g$.   The negative imaginary part of the QNFs confirms the stable modes of the obtained BH solution; However, at the large MM charge, the obtained BH solution is unstable.

%-------------------------------------------------------------------
\section{Results and Conclusions}\label{sec5}

In this work, we have constructed a new exact BH  solution when the gravity is minimally coupled to ABG NLED  and CS source. 
We have considered the ABG NLED term, which characterises the MM charge. 
The horizon structure of the obtained BH solution is explored. The size of the BH   decreases with an increase in MM charge, $g$, and increases with the CS parameter, $a$. The thermodynamics of BHs is also studied. Here, we have found that this BH solution follows the
modified first law of thermodynamics, which leads to entropy that follows the area law.

We numerically calculated the photon sphere radii and QNMs, including the shadow of the obtained BH solution. 
 We have found that the MM charge and CS parameter affect the photon radius and shadow radius.
In particular, the photon sphere radius and shadow radius increase with the CS parameter but decrease with the  MM  charge.  It will be attractive to 
investigate  the effects of thermal fluctuation 
  on the thermodynamics   of  
this BH.

\section*{Data Availability Statement} 
Data sharing does not apply to this article as no data sets were generated or analysed during the current study.

%%%%%%%%%%%%%%%%%%%%%%%%%%%%%%%%%%%%%%%%%%%%%%%%%%%%%%%%
\section*{Acknowledgement}
  DVS thanks the DST-SERB project (grant no. EEQ/2022/000824) and IUCAA, Pune, for the hospitality while this work was being done.

%%%%%%%%%%%%%%%%%%%%%%%%%%%

   \newpage
\appendix
\section{Ricci curvature invariants}
The   Ricci scalar ($R$) is  given by
\begin{eqnarray}
R&=&\frac{2a}{r^2}-\frac{24 g^2 r^2}{(r^2+g^2)^4}+ \frac{30 Mr^4}{(r^2+g^2)^{7/2}}+\frac{36 g^2 r^2}{(r^2+g^2)^3}-\frac{54M r^2}{(r^2+g^2)^{5/2}}-\frac{12 g^2}{(r^2+g^2)^2}\nonumber\\
&+&\frac{24M}{(r^2+g^2)^{3/2}}{.}
\label{rs}
\end{eqnarray}
Without the CS parameter and MM charge, the Ricci scalar (\ref{rs}) is zero.

The Ricci square ($R_{\mu\nu}  R^{\mu\nu}$) is  given by
\begin{eqnarray}
R_{\mu\nu}  R^{\mu\nu} &=&\frac{2a^2}{r^4}+\frac{288g^4r^8}{(r^2+g^2)^8}-\frac{720 g^2 M r^8}{(r^2+g^2)^{15/2}}-\frac{672 g^4 r^6}{(r^2+g^2)^7}+\frac{450M^2 r^8}{(r^2+g^2)^7}+\frac{1848 g^2M r^6}{(r^2+g^2)^{13/2}}\nonumber\\
&+&\frac{568 g^4r^4}{(r^2+g^2)^6}+\frac{1260 M^2r^6}{(r^2+g^2)^6}-\frac{1740g^2M r^4}{(r^2+g^2)^{11/2}}-\frac{216 g^4 r^2}{(r^2+g^2)^5}+\frac{1314M^2r^4}{(r^2+g^2)^5}+\frac{756 g^2 M r^2}{(r^2+g^2)^{9/2}}\nonumber\\&+&\frac{36 g^4}{(r^2+g^2)^4}-\frac{648 M^2r^2}{(r^2+g^2)^2}-\frac{144g^2 M}{(r^2+g^2)^{7/2}}+\frac{16ag^2}{(r^2+g^2)^3}+\frac{144M^2}{(r^2+g^2)^3}+\frac{144M^2}{(r^2+g^2)^3}\nonumber\\&-&\frac{24 a M}{(r^2+g^2)^{5/2}}-\frac{12 ag^2}{r^2(r^2+g^2)^2}+\frac{24 a M}{r^2(r^2+g^2)^{3/2}}{.}
\label{rs1}
\end{eqnarray}
Without the CS parameter and MM charge, the Ricci scalar (\ref{rs1}) is zero.

The Kretschmann 
scalar  ($R_{\mu\nu\lambda\sigma}R^{\mu\nu\lambda\sigma}$) is calculated by
\begin{eqnarray}
R_{\mu\nu\lambda\sigma}R^{\mu\nu\lambda\sigma}&=& \frac{4a^2}{r^4}+\frac{576g^4r^8}{(r^2+g^2)^8}-\frac{1440 g^2 M r^8}{(r^2+g^2)^{15/2}}-\frac{960 g^4 r^6}{(r^2+g^2)^7}+\frac{900M^2 r^8}{(r^2+g^2)^7}\nonumber\\&+&\frac{2640 g^2M r^6}{(r^2+g^2)^{13/2}}+\frac{560 g^4r^4}{(r^2+g^2)^6}+\frac{1800 M^2r^6}{(r^2+g^2)^6}-\frac{1704g^2M r^4}{(r^2+g^2)^{11/2}}-\frac{144 g^4 r^2}{(r^2+g^2)^5}
\nonumber\\
&+&\frac{1284 M^2r^4}{(r^2+g^2)^5}+\frac{504 g^2 M r^2}{(r^2+g^2)^{9/2}}+\frac{24 g^4}{(r^2+g^2)^4}-\frac{432 M^2r^2}{(r^2+g^2)^2}-\frac{96 g^2 M}{(r^2+g^2)^{7/2}}\nonumber\\
&+&\frac{96 M^2}{(r^2+g^2)^3}-\frac{8 ag^2}{r^2(r^2+g^2)^2}+\frac{16 a M}{r^2(r^2+g^2)^{3/2}}.
\label{rs3}
\end{eqnarray}
The Kretschmann 
scalars ($R_{\mu\nu\lambda\sigma}R^{\mu\nu\lambda\sigma}$) (\ref{rs3}) reduces to $48M^2/r^6$ in the limit of $a=g=0$. These invariants diverge in the limit of $r \rightarrow 0$   and, thus, signify a singular black hole solution. In the absence of a CS parameter, these invariants become singular.

\begin{thebibliography}{99}
\bibitem{1}B. P. Abbott et al. (LIGO Scientific, Virgo),   Phys.
Rev. Lett. 116, 061102 (2016).
\bibitem{2} K. Akiyama et al. (Event Horizon Telescope),   Astrophys. J. Lett. 875, L1-L6 (2019).
\bibitem{sg}K. Akiyama et al. (Event Horizon Telescope),  Astrophys. J. Lett. 930, L12 (2022).
\bibitem{ned}M. Born and L. Infeld,   Proc. R. Soc. Lond. 144,
425 (1934).
\bibitem{wit}N. Seiberg and E. Witten,   J. High Energy Phys. 09, 032 (1999).
\bibitem{sud01}P. Paul, S. Upadhyay and D. V. Singh, Eur. Phys. J. Plus 138, 566 (2023).
\bibitem{sud02} Y. Myrzakulov, K. Myrzakulov, S. Upadhyay and D. V. Singh, Int. J. Geom.
Meth. Mod. Phys. 20, 2350121 (2023); K. Esmakhanova, Y. Myrzakulov, G. Nugmanova and R. Myrzakulov,  Int. J. Theor. Phys. 51, 1204  (2012).
\bibitem{sud03}B. Pourhassan, M. Dehghani, S. Upadhyay, I. Sakalli and D. V. Singh,
Mod. Phys. Lett. A 37, 2250230 (2022).
 \bibitem{das}D. V. Singh, A. Shukla and S. Upadhyay, Annals of Physics 447, 169157 (2022).
\bibitem{cos1}  V. A. De Lorenci, R. Klippert, M. Novello and J. M. Salim,  Phys. Rev. D 65, 063501 (2002). 
\bibitem{cos2} M. Novello, S. E. P. Bergliaffa and J. Salim,   Phys. Rev. D 69, 127301 (2004). 
\bibitem{cos3} R. P. Mignani et al.,  Mon. Not. R. Astron. Soc. 465, 492 (2016). 
\bibitem{fr} E. S. Fradkin and A. A. Tseytlin,  Phys. Lett. B 163, 123 (1985).

\bibitem{ay1}E. Ay\'on-Beato and A. Garc\'ia,  Phys. Rev. Lett. 80, 5056 (1998).
\bibitem{ay2} E. Ay\'on-Beato and A. Garc\'ia, Phys. Lett. B 464, 25 (1999).  
\bibitem{ay3}E. Ay\'on-Beato and A. Garc\'ia, Phys. Lett. B 493, 149 (2000).  
\bibitem{ay4} E. Ay\'on-Beato and A. Garc\'ia,  Gen. Rel. Grav. 37, 635 (2005). 
 \bibitem{Singh:2022xgi} D.~V.~Singh, S.~G.~Ghosh and S.~D.~Maharaj,
Nucl. Phys. B {981}, 115854  (2022).
\bibitem{Singh:2017qur}
D.~V.~Singh and N.~K.~Singh,
%``Anti-Evaporation of Bardeen de-Sitter Black Holes,''
Annals Phys.  {383},  600  (2017).
 \bibitem{Filho:2023voz}
A.~A.~A.~Filho, 
Class. Quant. Grav. {41}, 015003  (2024).

\bibitem{Biswas:2022qyl}
A.~Biswas, 
Gen. Rel. Grav. {54}, 161 (2022).

\bibitem{KumarWalia:2022aop}
R.~Kumar Walia, S.~G.~Ghosh and S.~D.~Maharaj, 
Astrophys. J. {939}, 77 (2022).

\bibitem{Belhaj:2022qmn}
A.~Belhaj and Y.~Sekhmani, 
Eur. Phys. J. Plus {137}, 278 (2022).
\bibitem{Kumar:2023cmo}
A.~Kumar, D.~V.~Singh, Y.~Myrzakulov, G.~Yergaliyeva and S.~Upadhyay, 
Eur. Phys. J. Plus {138}, 1071  (2023).
\bibitem{xu19}
 P. S. Letelier, Phys. Rev. D  {28}, 2414 (1983).
 \bibitem{lt1}P. S. Letelier, Il Nuovo Cim. B {63}, 519  (1981).
 \bibitem{lt2}
P.S. Letelier, Phys. Rev. D {20}, 1294 (1979).
\bibitem{Singh:2020nwo}
D.~V.~Singh, S.~G.~Ghosh and S.~D.~Maharaj, 
Phys. Dark Univ. {30}, 100730  (2020).



\bibitem{roh} R. Takahashi and  J. Korean
Phys. Soc. 45, S1808 (2004). 
\bibitem{ped} P. V. P. Cunha, C. A. R. Herdeiro  and M. J. Rodriguez,  Phys. Rev. D 97, 084020 (2018).
 \bibitem{Vishvakarma:2023tnl}
B.~K.~Vishvakarma, D.~V.~Singh and S.~Siwach, 
Phys. Scripta  {99},   025022 (2024).
\bibitem{Vishvakarma:2023csw}
B.~K.~Vishvakarma, D.~V.~Singh and S.~Siwach, 
Eur. Phys. J. Plus {138}, 536  (2023).
  

\bibitem{Zubair:2023cep}
M.~Zubair, M.~A.~Raza, F.~Sarikulov and J.~Rayimbaev,
JCAP {10}, 058  (2023).



\bibitem{ss}S. Mandal, S. Upadhyay, Y. Myrzakulov, G. Yergaliyeva, Int. J. Mod.
Phys. A 38, 2350047 (2023).
\bibitem{ss1} S. Upadhyay, S. Mandal, Y. Myrzakulov, and K. Myrzakulov, Annals of
Physics {450}, 169242 (2023).
\bibitem{ss2}D. V. Singh, V. K. Bhardwaj, S. Upadhyay, Eur. Phys. J. Plus 137, 969
(2022).
 
\bibitem{xx}J. Jing, Q. Pan, Phys. Lett. B {660}, 13 (2008) .
\bibitem{qnm}S. Chandrasekhar, S. Detweller, Proc. R. Soc. Lond. Ser. A Math. Phys. Eng. Sci. {344},  441 (1975).
\bibitem{Mou22}
P.-Hui Mou, Y.-Xian Chen, K.-Jian He and G.-Ping Li, Commun. Theor. Phys. {74},  125401  (2022).
\bibitem{Cai:2021ele}
X.~C.~Cai and Y.~G.~Miao, 
Phys. Rev. D {103}, 124050 (2021).

\bibitem{Gogoi:2021cbp}
D.~J.~Gogoi, R.~Karmakar and U.~D.~Goswami, 
Int. J. Geom. Meth. Mod. Phys. {20}, 2350007  (2023).

\bibitem{Murshid:2022ssj}
M.~Murshid, F.~Rahaman and M.~Kalam, 
Indian J. Phys. {97}, 295   (2023).

 \bibitem{Konoplya:2023ahd}
R.~A.~Konoplya, D.~Ovchinnikov and B.~Ahmedov, 
Phys. Rev. D {108}, 104054 (2023).
\bibitem{Gogoi:2023ffh}
D.~J.~Gogoi, J.~Bora, M.~Koussour and Y.~Sekhmani, 
Annals Phys. {458}, 169447 (2023).

\bibitem{Jafarzade:2021umv}
K.~Jafarzade, M.~Kord Zangeneh and F.~S.~N.~Lobo, 
Annals Phys. {446},  169126 (2022).

\bibitem{Ladino:2023zqc}
J.~M.~Ladino and E.~Larra\~naga, 
Int. J. Theor. Phys. {62}, 209 (2023).
\bibitem{Li21}
P.-C. Li, T.-C. Lee, M. Guo, and B. Chen,
 Phys. Rev. D {104}, 084044 (2021).
\bibitem{k20}
 K. Jusufi, 
Phys. Rev. D {101}, 124063 (2020).
\bibitem{11}J. D. Bekenstein,  Lett. Nuovo Cim.  {4},  737 (1972).
 
\bibitem{bak} J. D. Bekenstein,  Phys. Rev. D  {7},  2333 (1973).
\bibitem{bak0} S. W. Hawking,   Phys. Rev. D  {13}, 191 (1976).

\bibitem{str} A. Strominger and C. Vafa,   Phys. Lett. B  {379}, 99  (1996).
 \bibitem{car} S. Carlip,  Phys. Rev. Lett.  {82}, 2828  (1999). 

 
\bibitem{sud2} S. Upadhyay,   Phys. Lett. B  {775}, 130 (2017).
\bibitem{sud3} S. Upadhyay, B. Pourhassan and H. Farahani,     Phys. Rev. D  {95}, 106014 (2017). 
\bibitem{jy}S. Upadhyay, N.-ul-Islam, P. A. Ganai, JHAP 2, 25 (2022).
\bibitem{bss}  B. Pourhassan, H. Farahani, F. Kazemian, Izzet Sakalli, S. Upadhyay,
D. V. Singh, Physics of the Dark Universe 44, 101444 (2024).
\bibitem{sud1} B. Pourhassan, S. Upadhyay, H. Saadat and H. Farahani,  Nucl. Phys. B 928, 415 (2018).
\bibitem{behn}S. Upadhyay,   Gen. Rel. Grav.   {50}, 128 (2018).
\bibitem{jhap}S. Chougule, S. Upadhyay, H. K. Sudhanshu and S. Kumar, JHAP 3, 45
(2023).
\bibitem{jhap1}S. S. Shahraeini, K. Nozari and S. Saghafi, JHAP 2,    55 (2022).
\bibitem{beh1}S. Soroushfar, B. Pourhassan and I. Sakalli, Physics of the Dark Universe 44, 101457 (2024).
\bibitem{beh2} B. Pourhassan, I. Sakalli, X. Shi, M. Faizal and S. S. Wani, EPL 144, 29001 (2023).
\bibitem{beh3}B. Pourhassan and I. Sakalli, Chinese Journal of Physics 79, 322 (2022).
\bibitem{sud7}M. Yasir, T. Xia and S. Upadhyay, Phys. Scr. 99, 065003 (2024).
 
 \bibitem{24}
 J. L. Synge, Relativity: The General Theory,  p. 175v,(North Holland, Amsterdam, 1966).
 \bibitem{bar}
M. Barriola and A. Vilenkin, Phys. Rev. Lett. {63}, 341 (1989).
\bibitem{kibb76}
 T. W. B. Kibble, J. Phys. A  {9}, 1387 (1976).
 %%%%%%%%%%%%%%%%%%%%%%%%%%%%%%%%%%

\bibitem{uma}
S.~G.~Ghosh, U.~Papnoi and S.~D.~Maharaj, 
Phys. Rev. D  {90},  044068 (2014).
\bibitem{Singh:2021nvm}
B.~K.~Singh, R.~P.~Singh and D.~V.~Singh, 
Eur. Phys. J. Plus {136},    575 (2021).
 %%%%%%%%%%%%%%%%%%%%%%%%%%%%%%%%%%
 \bibitem{ma14}
M.-Sen Ma and R. Zhao, Class. Quant. Grav. {31},  245014 (2014).
\bibitem{dharm}
S.~G.~Ghosh, D.~V.~Singh and S.~D.~Maharaj, 
Phys. Rev. D {97}, 104050 (2018).
\bibitem{page}
S. W. Hawking and  D. N. Page, Commun. Math. Phys. 87, 577  (1983).
   \bibitem{jh87}
J. H. MacGibbon, Nature (London) {329}, 308 (1987).
  \bibitem{p92}
 J. Preskill, 	arXiv: hep-th/9209058.
  
\bibitem{Belhaj:2020okh}
A.~Belhaj, H.~Belmahi, M.~Benali, W.~El Hadri, H.~El Moumni and E.~Torrente-Lujan, 
Phys. Lett. B  {812},  136025 (2021) .

\bibitem{Belhaj:2022kek}
A.~Belhaj and Y.~Sekhmani, 
Gen. Rel. Grav. {54},   17 (2022).

\bibitem{Belhaj:2021tfc}
A.~Belhaj, H.~Belmahi, M.~Benali and A.~Segui,
Phys. Lett. B  {817}, 136313 (2021).
  \bibitem{72}V.~Perlick, O.~Y.~Tsupko and G.~S.~Bisnovatyi-Kogan, 
Phys. Rev. D  {92}, 104031 (2015).
\bibitem{will}
B. F. Schutz and C. M. Will, Astrophys J. {291}, L33 (1985).
\bibitem{iyer}
S. Iyer and C. M. Will, Phys. Rev. D {35}, 3621(1987).
\bibitem{konoplya1}
R. A. Konoplya, Phys. Rev. D {68}, 024018 (2003).

\bibitem{wkb1}
V.~Cardoso, A.~S.~Miranda, E.~Berti, H.~Witek and V.~T.~Zanchin, 
Phys. Rev. D  {79}, 064016 (2009).

%%%%%%%%%%%%%%%%%%%%%%%%%%%%%%%%%%%%%%%%%%%%








   \end{thebibliography}
\end{document}